\documentclass[aps,twocolumn,letterpaper,longbibliography]{revtex4-1}
\usepackage{amssymb}
\usepackage{amsmath}
\usepackage{graphicx}
\usepackage{extarrows}
\usepackage{url}
\usepackage{varioref}
\usepackage{hyperref}
\usepackage{cleveref}

\newcommand{\vk}{\mathbf{k}}

\begin{document}
\title{Complex correlations in high harmonic generation of matter-wave jets revealed by pattern recognition}
\author{Lei Feng}
\author{Jiazhong Hu}
\author{Logan W. Clark}
\author{Cheng Chin}
\affiliation{James Franck Institute, Enrico Fermi Institute and Department of Physics, the University of Chicago, Chicago, IL, 60637, USA}

\begin{abstract}

Correlations in interacting many-body systems are key to the study of quantum materials and quantum information. More often than not, the complexity of the correlations grows quickly as the system evolves and thus presents a challenge for experimental characterization and intuitive understanding. In a strongly driven Bose-Einstein condensate, we observe the high harmonic generation of matter-wave jets with complex correlations as a result of bosonic stimulation. Based on a pattern recognition scheme, we identify a universal pattern of correlations which offers essential clues to unveiling the underlying secondary scattering processes and high-order correlations. We show that the pattern recognition offers a versatile strategy to visualize and analyze the quantum dynamics of a many-body system.

\end{abstract}

\maketitle

High harmonic generation is an elegant phenomenon in nonlinear optics, which transfers photon populations to specific excited modes, and enables modern applications such as X-ray sources \cite{Chang1997}, attosecond spectroscopy \cite{Hentschel2001,Drescher2001} and frequency combs \cite{JunYe2005,Gohle2005}. The generation of high harmonics relies on both the nonlinearity of the coupling between photons and particles and a strong coherent driving \cite{Boyd2008}.

Atom optics, the matter-wave analog of optics, naturally inherits nonlinearity from atomic interactions \cite{Phillips2002, AtomOpticsRev}. The matter-wave versions of lasers \cite{Mewes1997,Anderson1998,Munich1999,Hagley1999}, super-radiance \cite{Inouye571,Moore1999,Schneble2003}, four-wave mixing \cite{Deng1999, Julienne2000}, Faraday instability \cite{FDwaveT,FDwaveE} and spin-squeezing \cite{Gross2010,Riedel2010}, have made manifest the quantum coherence of matter waves.
In particular, experimental characterization of high-order correlations \cite{Guarrera2011, Dall2013,Schweigler2017} has been demonstrated  by splitting and interfering a condensate \cite{Andrews1997,Hall1998}, akin to homodyne detection with lasers. 
Beyond analogies to quantum optics, the manipulation of coherent matter waves can also offer a unique platform to simulate large-scale \cite{Hung1213,Steinhauer2016,Feng2017} and high-energy physics \cite{Arratia2018}.

In this work, we demonstrate high harmonic generation of matter waves by strongly modulating the interactions between atoms in a Bose condensate. Matter waves emerging from the driven condensates form jet-like emission (Bose fireworks) \cite{jets}. Above a threshold in the driving amplitude, a quantized spectrum of the matter-wave manifests due to bosonic stimulation, whose temporal evolution suggests a hierarchy of the atomic emission process. By applying a pattern recognition algorithm \cite{Flusser1993, nas1}, we identify intriguing second- and higher-order correlations of emitted atoms that are not obvious from individual experiments. Our machine learning strategy provides new prospects for analyzing complex dynamical systems.

\begin{figure}[tb]
\begin{center}
\includegraphics[width=87mm]{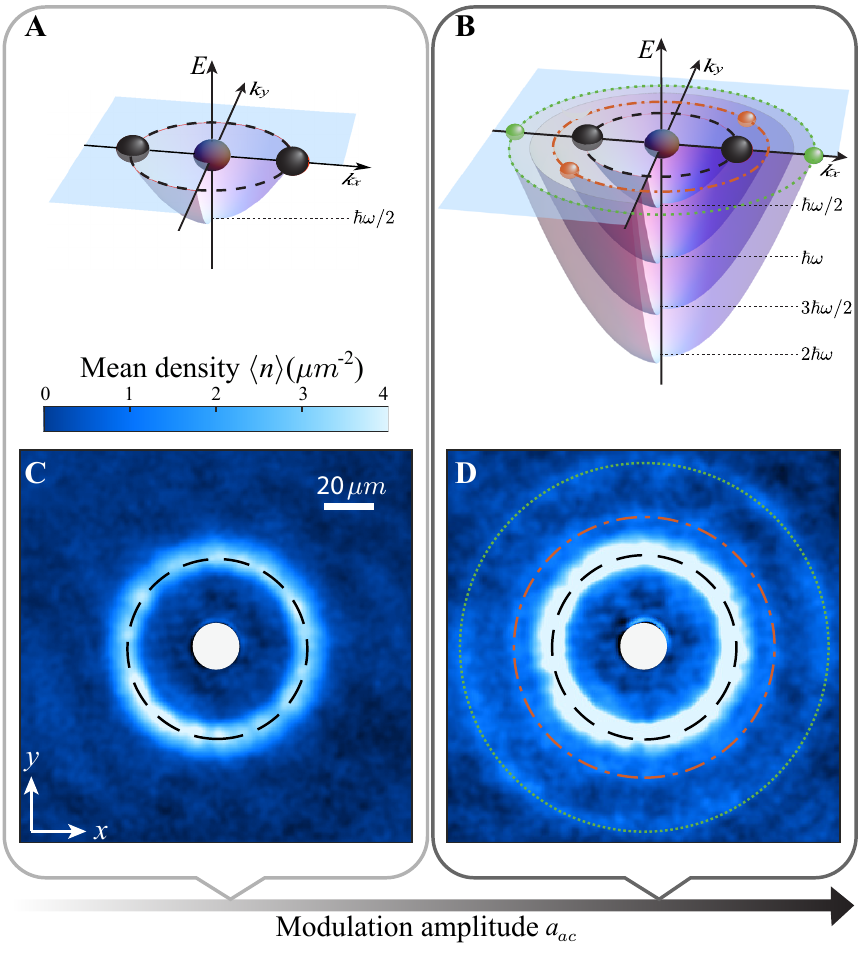}
\caption{The first and high harmonic generation of matter-wave jets in driven condensates. Figures \textbf{A} and \textbf{B} show the dispersion relation between energy $E$ and momentum $\hbar\vk=\hbar\left(k_x,k_y\right)$ in the dressed-state picture. \textbf{C} shows the average \textit{in situ} image of emitted atoms at a small modulation amplitude $a_{ac}$~=~25~$a_0$. The emission pattern displays a single ring (black circle) indicating the generation of matter-wave jets with momentum $k_f$. \textbf{D} shows two more rings (orange and green circles) in the average image at a larger modulation amplitude $a_{ac}$~=~45~$a_0$.  Atoms in these three rings have quantized kinetic energy of $\hbar \omega/2$, $\hbar \omega$, and $2\hbar \omega$ respectively. The \textit{in situ} images are taken at $21$~ms after the beginning of the modulation.}
\label{fig1}
\end{center}
\end{figure}

The experiment starts with Bose-Einstein condensates of $6\times 10^4$ cesium atoms loaded into a uniform disk-shaped trap with a radius of 7~$\mu$m, a barrier height of $h\times$300~Hz in the horizontal direction and harmonic trapping frequency of 220~Hz in the vertical direction \cite{jets}, where $h=2\pi\hbar$ is the Planck constant. The interaction between atoms, characterized by the $s$-wave scattering length $a$, can be tuned near a Feshbach resonance by varying the magnetic field \cite{RevModPhys.82.1225}.

After the preparation, we oscillate the scattering length as $a(t)=a_{dc}+a_{ac}\sin(\omega t)$ for a short period of time $\tau$=~5~ms with a small DC value $a_{dc}$~=~3~$a_0$ and a tunable amplitude $a_{ac}$ at frequency $\omega=2\pi\times 2$~kHz.  Here $a_0$ is the Bohr radius. We then perform either \textit{in situ} imaging or time-of-flight measurement on the sample. At a modulation amplitude $a_{ac}$~=~$25$~$a_0$ we see the dominant emission of matter-wave jets with each atom emitted at kinetic energy of $\hbar\omega/2$, as evidenced by the velocity with which they leave the sample \cite{jets}.The angles of the emitted jets vary randomly from shot to shot, resulting in a single isotropic ring of atoms after averaging images from many trials (Fig.~\ref{fig1}\textbf{A}). At a larger amplitude $a_{ac}$~=~$45$~$a_0$ multiple rings form, labeled as ring 1, ring 2 and ring 4 (Fig.~\ref{fig1}\textbf{B}). Atoms in each ring have quantized kinetic energy of $E_j=j\hbar\omega/2=j\hbar^2k_f^2/2m$ with $j=1$, 2, and 4, where $k_f=\sqrt{m\omega/\hbar}$ is the characteristic wave number of the jets and $m$ is the atomic mass.

To describe atoms with oscillating scattering length, we write down the Hamiltonian of the system  \cite{SM}
\begin{eqnarray}
H&=&\sum_{\mathbf{k}} \epsilon_{\mathbf{k}} a^\dagger_{\mathbf{k}} a_{\mathbf{k}}+\hbar\omega b^\dagger b \nonumber \\
& &+\sum_{\mathbf{k_1},\mathbf{k_2},\Delta \mathbf{k}}\left(Aa^\dagger_{\mathbf{k_1}+\Delta \mathbf{k}} a^\dagger_{\mathbf{k_2}-\Delta \mathbf{k}} a_{\mathbf{k_1}} a_{\mathbf{k_2}}b+h.c.\right), \label{meq1}
\end{eqnarray}
where $\epsilon_{\mathbf{k}}=\hbar^2\mathbf{k}^2/2m$ is the kinetic energy of free atoms, $a_\mathbf{k}$ ($a^\dagger_\mathbf{k}$) is  the annihilation (creation) operator of an atom with momentum $\hbar \mathbf{k}$, $b$ ($b^\dagger$) is the  annihilation (creation) operator of photon with energy of $\hbar\omega$ associated with the magnetic field modulation, and $A$ is the coupling strength between atoms and the field. The resonant terms that satisfy energy conservation, 
\begin{equation}
\epsilon_{\mathbf{k_1}+\Delta \mathbf{k}}+\epsilon_{\mathbf{k_2}-\Delta \mathbf{k}}=\epsilon_{\mathbf{k_1}}+\epsilon_{\mathbf{k_2}}\pm\hbar\omega, 
\label{eq:conservation}
\end{equation}
describe the dominant collision processes. The momentum and recoil energy of the photon in our experiment are negligible. 

The Hamiltonian describes a five-wave mixing process where, by absorbing or emitting one photon, an atom pair increases or decreases its total kinetic energy by an energy quantum $\hbar\omega$. Based merely on the conservation of energy and momentum, the five-wave mixing can produce atoms in a continuous spectrum of energy states. However, given bosonic stimulation, we expect a quantized energy spectrum of the emitted atoms. Here starting with the condensate, the collisions first excite atoms to ring 1. As the population in ring 1 builds up, atoms can be further promoted to higher momentum modes through the matter-wave mixing of the condensate and the atoms in ring 1.  Because of bosonic stimulation, such process is dominated by scattering involving three macroscopically occupied modes and the fourth unoccupied mode with higher energy. Therefore a hierarchy of stimulated collisions is expected.
From energy conservation Eq.~(\ref{eq:conservation}), atoms in the fourth mode acquire discrete energies $E_j=j\hbar\omega/2$ with $j=2,3,4,...$, analogous to the photon spectra from high harmonic generation.

\begin{figure}[t]
\begin{center}
\includegraphics[width=56mm]{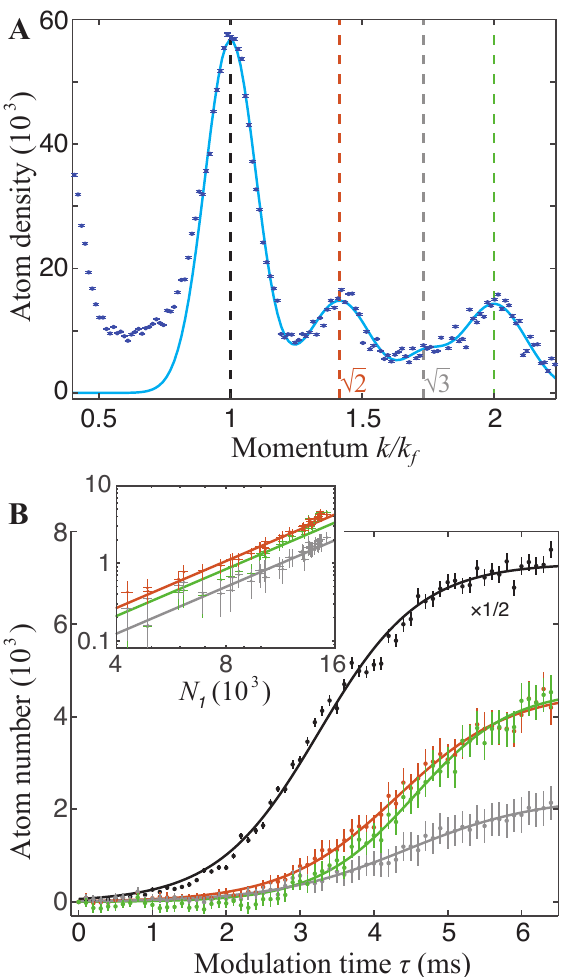}
\caption{The atomic population growth in multiple rings. \textbf{A} shows a snapshot of the population distribution in momentum space measured by the focused-TOF imaging $\tau$~=~6~ms after the modulation starts.  The cyan curve is a fit to the experimental data for $k/k_f$~$>$~0.85 using a combination of 4 Gaussians. The vertical dashed lines indicate centers of the Gaussians, fixed at $k/k_f$~=~1 (black), $\sqrt{2}$ (orange), $\sqrt{3}$ (gray), and 2 (green), respectively. \textbf{B} shows the extracted atom number in each ring as a function of modulation time $\tau$. The atom number in ring 1 (black) is scaled by a factor of 1/2. Populations in ring 2 (orange), 3 (gray), and 4 (green) arise after ring 1 is significantly populated. The inset shows the atom number in ring 2, 3 and 4 as a function of atom number in ring 1. The solid lines are the power-law fits to the data with the exponent fixed to 2. The error bars represent one standard error.}
\label{fig2}
\end{center}
\end{figure}

To verify this picture, we inspect the evolution of atomic population in each ring using time-of-flight imaging \cite{Feng2017, SM}. Figure~\ref{fig2}\textbf{A} shows an example of the momentum distribution after modulating the scattering length for $\tau$~=~5~ms. Beside the distinct peaks at $|\mathbf{k}|$ = $k_f$, $\sqrt{2}k_f$ and $2k_f$, which are apparent from \textit{in situ} images (Fig.~\ref{fig1}), we also detect a much weaker peak at $|\mathbf{k}|$ = $\sqrt{3}k_f$ (ring 3). We fit the density distribution using a combination of four Gaussians with fixed central positions and widths to extract the population $N_j$ in ring $j$.

Populations in all four rings initially show a fast exponential growth and gradually saturate afterward (Fig.~\ref{fig2}\textbf{B}). The population growth of rings 2, 3 and 4 are delayed from that of ring 1. Furthermore, we observe that the population in high-order rings are proportional to the square of that in ring 1, $N_j\propto N_1^2$ with $j$ = 2, 3 and 4 (Fig.~\ref{fig2}\textbf{B} inset). Since the population grows exponentially, this relation is equivalent to $\dot{N_j}\propto N_1^2$, which suggests that the production of atoms in the these rings involves two modes in ring 1, which is in agreement with our model \cite{SM}. We thus consider these processes as secondary collisions, which occur after ring 1 is populated by primary collisions.

\begin{figure*}[t]
\begin{center}
\includegraphics[width=116mm]{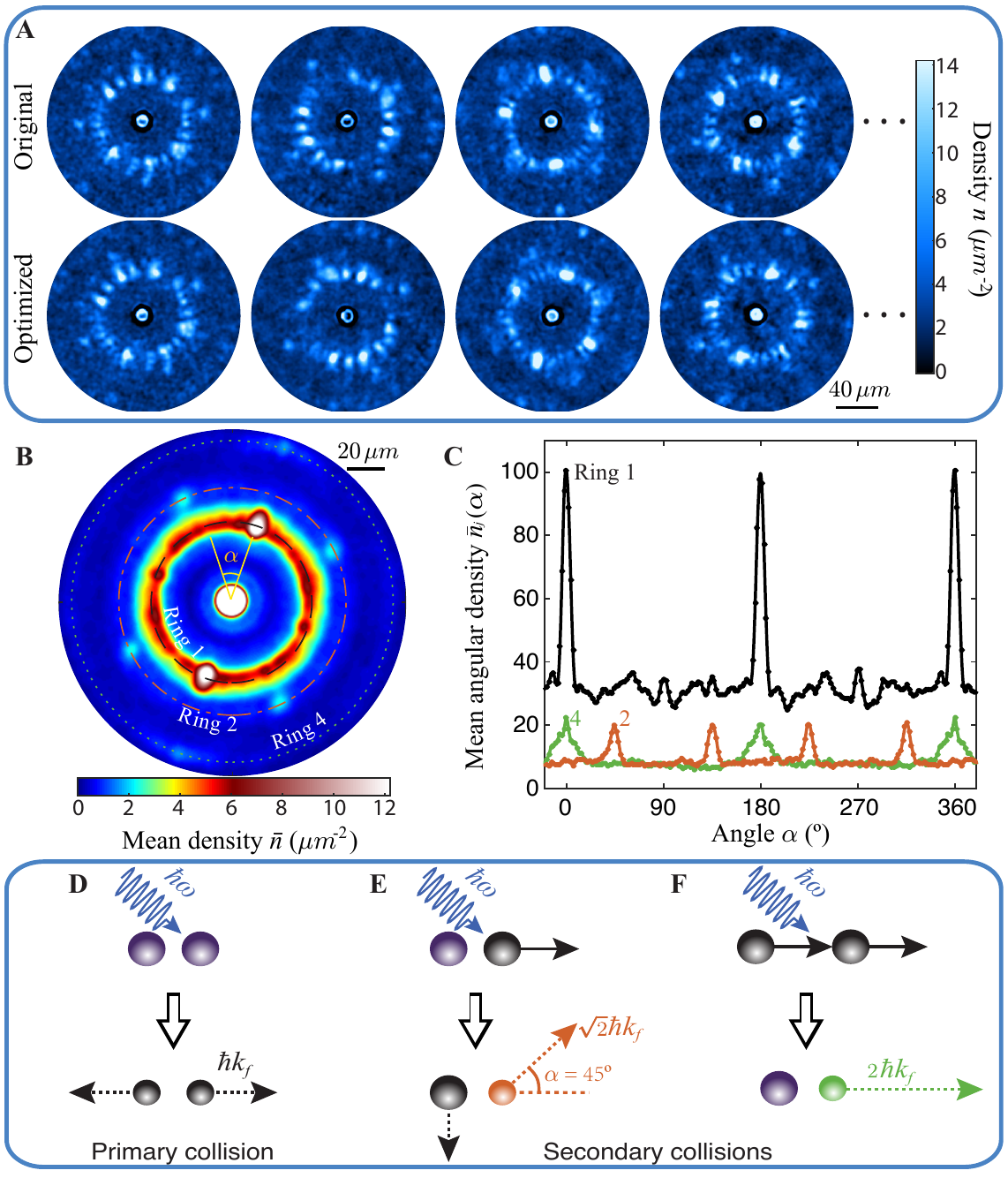}
\caption{Pattern recognition and microscopic interpretation.
 \textbf{A} shows examples from a dataset of 209 raw images (top) and those after individual rotation (bottom) that maximized the angular variance of the mean image.
\textbf{B} shows the resulting pattern $\mathbf{\Phi}$ from pattern recognization, namely, the average of all 209 images after individual rotation. Besides the bright center spot corresponding to the remnant condensate, ten more distinct spots emerge in an angular-uniform background: four in ring 1 at $\alpha$~=~0$^\circ$, 90$^\circ$, 180$^\circ$ and 270$^\circ$; four in ring 2 at $\alpha$~=~45$^\circ$, 135$^\circ$, 225$^\circ$ and 315$^\circ$; and two in ring 4 at $\alpha$~=~0$^\circ$ and 180$^\circ$. \textbf{C} shows the angular density distributions in ring 1, 2 and 4 extracted from the pattern, where the ten bright spots in the pattern show up as peaks in the distributions.
\textbf{D}, \textbf{E} and \textbf{F} illustrate the microscopic processes that are responsible for the peaks. The purple balls indicate atoms in the condensate. The black, orange and green balls represent atoms in ring 1, 2 and 4 with momentum of $k_f$, $\sqrt{2}k_f$, and $2k_f$, respectively.}
\label{fig3}
\end{center}
\end{figure*}

Beyond the population growth, emissions from secondary collisions display a wealth of intriguing angular structures (Fig.~\ref{fig3}\textbf{A}) that are not obvious from the Hamiltonian in Eq.~(\ref{meq1}) or the average image. To investigate these structures, we employ a pattern-recognition algorithm based on unsupervised machine learning. Here we collect and analyze 209 independent images taken under the same conditions as those in Fig.\ref{fig1}\textbf{B}. We rotate each image $I_i$ with an angle of $\theta_i$ around the center of the condensate and maximize the angular variance of the mean image $\bar{I}$ by tuning all 209 angles \cite{SM}.

The algorithm recognizes a robust and intriguing pattern in the jet emission, defined as the pattern $\mathbf{\Phi}$, containing multiple distinct spots at non-zero momenta on top of angularly uniform rings 1, 2 and 4(Fig.~\ref{fig3}\textbf{B}). To better characterize these features, we extract the mean angular density $\bar{n}_j(\alpha)$ for each ring with $\alpha$ the relative angle to the brightest spot in ring 1 (Fig.~\ref{fig3}\textbf{C}). In this way, we convert the pattern into a series of angular density plots with a flat background and clear peaks representing the spots. This flat background contains a combination of emissions of atoms uncorrelated to the main pattern $\mathbf{\Phi}$. Note that any possible contribution from ring 3 is too weak to discern; for the remainder of this work, we focus on the stronger signals from rings 1, 2, and 4.

Excluding the uniform background, we find that the concurrence of multiple spots in the pattern points to particular scattering processes populating the corresponding momentum modes. As the first example, two strong peaks in ring 1 ($\alpha$~=~0$^\circ$ and 180$^\circ$) come from primary collisions of two condensate atoms, which absorb one energy quantum and are scattered into opposite directions with momentum $\pm \hbar k_f$, shown in Fig.~\ref{fig3}\textbf{D}.

Following the primary collisions, stimulated secondary collisions induce eight additional peaks in total among the three rings. We consider that the four peaks in ring 2 (at $\alpha$~=~45$^\circ$, 135$^\circ$, 225$^\circ$ and 315$^\circ$) and two peaks in ring 1 (at $\alpha$~=~90$^\circ$ and 270$^\circ$) arise from the collisions between an atom from ring 1 and another atom from the condensate. One example of such collisions is illustrated in Fig.~\ref{fig3}\textbf{E}, where a pair of atoms populate two specific modes at $\alpha$~=~45$^\circ$ in ring 2 and at $\alpha$~=~270$^\circ$ in ring 1 by absorbing one photon.
Another secondary collision process, shown in Fig.~\ref{fig3}\textbf{F}, can explain the origin of the two peaks in ring 4. Here two co-propagating atoms from ring 1 collide; one atom is promoted to ring 4  and the other returns to the condensate. The above two scattering processes satisfying energy-momentum conservation are matter-wave analogs to four-wave mixing with phase matching condition. These processes are the dominant secondary collisions because they involve as many macroscopically occupied momentum modes as possible \cite{SM}.

To further support the dominant microscopic collision processes implied by the pattern $\mathbf{\Phi}$, we calculate the second-order correlation function $g^{(2)}_{ij}(\phi)$ between momentum modes in ring $i$ and ring $j$, namely,
\begin{equation}
g^{(2)}_{ij}(\phi)={\langle n_i(\theta)[n_j(\theta+\phi)-\delta_{ij}\delta(\phi)]\rangle\over\langle n_i(\theta) \rangle\langle n_j(\theta+\phi) \rangle },
\end{equation}
where $n_i(\theta)$ is the angular density in ring $i$ at angle $\theta$, $\delta_{ij}$ is the Kronecker delta, and $\delta(\phi)$ is the Dirac delta function. The angle brackets correspond to angular averaging over $\theta$, followed by ensemble averaging over images.

\begin{figure*}[htbp]
\begin{center}
\includegraphics[width=116mm]{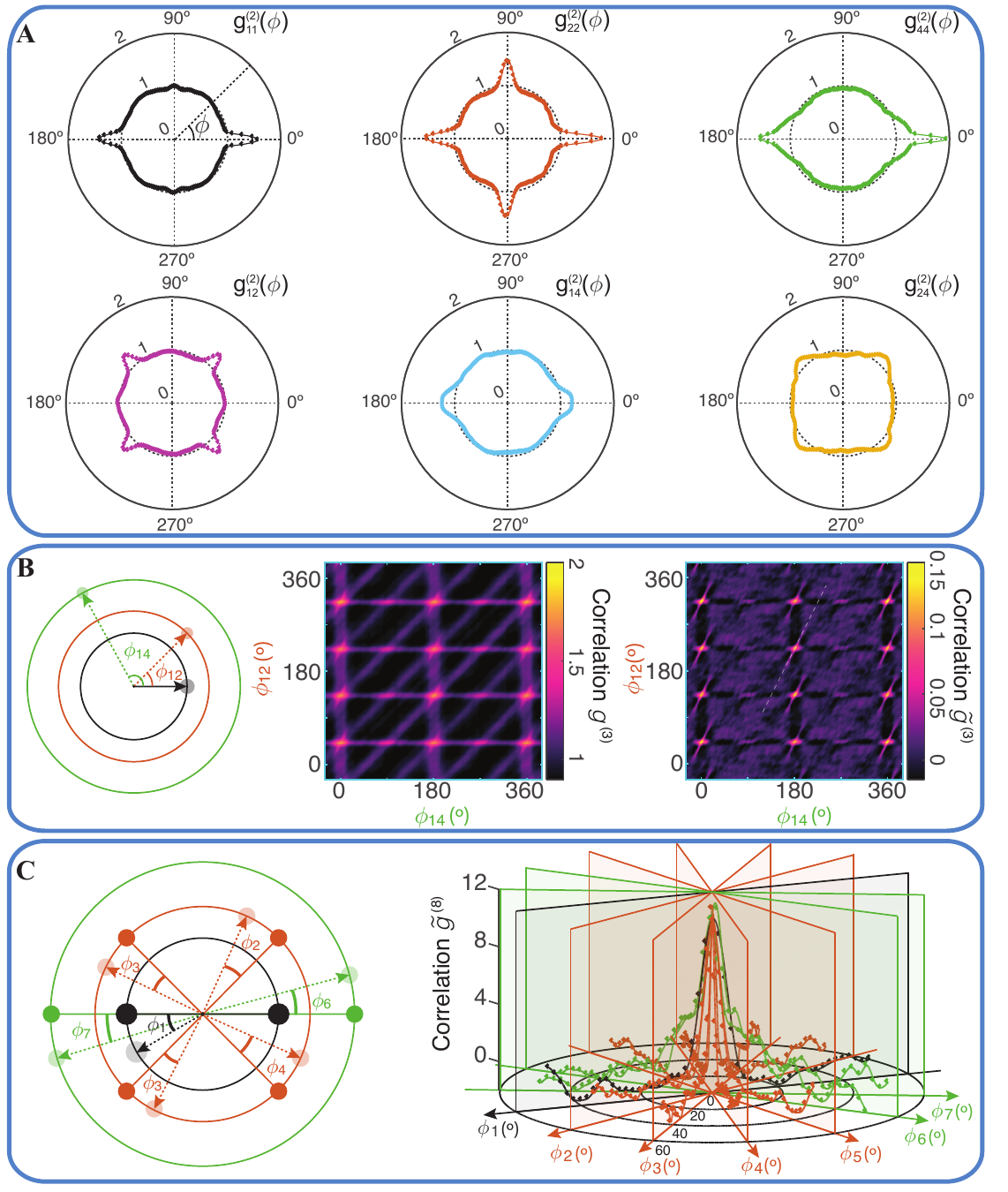}
\caption{Second-, third- and eighth-order correlations of emitted matter-wave jets.
\textbf{A} shows all the second-order correlation functions $g_{ij}^{(2)}(\phi)$ within and between rings.
\textbf{B} shows the third-order correlations $g_{124}^{(3)}(\phi_{12} , \phi_{14})$ and the connected part $\tilde{g}_{124}^{(3)}(\phi_{12} , \phi_{14} )$. Here $\phi_{12}$ ($\phi_{14}$ ) are the relative angles between atoms in ring 1 and 2 (4), shown in the left figure. The extended weak correlations across the peaks in $\tilde{g}_{124}^{(3)}(\phi_{12} , \phi_{14} )$ indicate a relation between small angular deviations $\delta\phi_{12}=2\delta\phi_{14}$ (see white-dashed line as an example).
\textbf{C} shows the eighth-order connected correlation function $\tilde{g}^{(8)}(\phi_1,...,\phi_7)$ \cite{SM}. We choose a primary jet direction in ring 1 as the reference and show the connected correlation as a function of seven angles relative to the locations of seven bright spots in pattern $\mathbf{\Phi}$: one in ring 1 (at 180$^\circ$, black), four in ring 2 (at 45$^\circ$, 135$^\circ$, 225$^\circ$ and 315$^\circ$, orange), and two in ring 4 (at 0$^\circ$ and 180$^\circ$, green). The correlation is shown on seven vertical planes. Within each plane, only one of the seven angles is varied. The connected $\tilde g^{(8)}$ reaches 12 and decays rapidly to zero as the angles increase, suggesting a peak in the 7-dimensional space. Here solid lines are spline fits to guide the eye.
}
\label{fig4}
\end{center}
\end{figure*}

All of the second-order correlations involving momenta on the dominant rings display multiple peaks (Fig.~\ref{fig4}\textbf{A}). The results are in full consistency with the spots in the pattern $\mathbf{\Phi}$ and the collisional processes that we identify. In particular, we can associate all the peaks in $g^{(2)}_{22}$ and $g^{(2)}_{12}$ with the process shown in Fig.~\ref{fig3}\textbf{E}, where jets in ring 2 are created at $\pm$45$^\circ$ relative to the primary jets.  The peaks in $g^{(2)}_{44}$ and $g^{(2)}_{14}$ are associated with the process in Fig.~\ref{fig3}\textbf{F}, where jets are created along the direction of the primary jets.

We also find four peaks in the cross-correlation between rings 2 and 4. To the best of our knowledge, these correlations cannot come directly from a single secondary collision process.  Instead, they could result from the concurrence of two secondary collision processes. Such correlation develops since both processes involve the same macroscopically-occupied modes in ring 1 and the condensate. 

We further investigate such indirect correlation by calculating the third-order correlation function $g_{124}^{(3)}(\phi_{12},\phi_{14})$ between ring 1, 2 and 4 (Fig.~\ref{fig4}\textbf{B} top right), where $\phi_{ij}$ is the relative angle between emitted atoms in ring $i$ and ring $j$.  To remove contributions from the lower-order correlations, we evaluate the connected correlation function $\tilde g^{(3)}$ defined as \cite{SM,Schweigler2017}
\begin{eqnarray}
\tilde g_{124}^{(3)}=g_{124}^{(3)}-g_{12}^{(2)}(\phi_{12})- g_{14}^{(2)}(\phi_{14})-g_{24}^{(2)}(\phi_{24})+2.
\end{eqnarray}
The results are shown in Fig.~\ref{fig4}\textbf{B} (bottom right). 

The distinct peaks in the connected third-order correlation reveals genuine bunching of the population fluctuations at specific angles in all three rings. In addition, we observe extended weak correlations along lines across the peaks, which relate the angular deviations $\delta\phi_{12}\approx 2\delta\phi_{14}$ (see Fig.\ref{fig4}\textbf{B}). We attribute such weak correlations to a Raman-like collision process that couples these three momentum modes by absorbing two energy quanta from the modulation field \cite{SM}. 

Beyond third-order correlations, high harmonic generation can induce even higher order correlations. An example shown in Fig.~\ref{fig4}\textbf{C} is the connected eighth-order correlation. Plotted in the seven-dimensional space spanned by the angular deviations, a prominent peak appears when the angles match the bright spots in our pattern $\mathbf{\Phi}$. 

In conclusion, we demonstrate the high harmonic generation of matter-wave jets with quantized energy and momentum from strongly driven Bose condensates. With the assistance of pattern recognition, mutual correlations of matter-wave jets are visualized, which guides us to discover the underlying dominant secondary and higher-order collision processes. 

Our experiment shows that bosonic stimulation in a driven system can connect different momentum modes in a coherent manner. This suggests a novel way to prepare highly correlated systems for applications in quantum simulation and quantum information. In addition, the implementation of the pattern recognition can inspire further applications of machine learning to understand complex dynamics of quantum systems. 

We thank B. M. Anderson, K. Levin, Z. Zhang and M. McDonald for helpful discussions. L. F. is supported by MRSEC Graduate Fellowship. L. W. C.  was supported by Grainger Graduate Fellowship. This work is supported by the University of Chicago Materials Research Science and Engineering Center, which is funded by the National Science Foundation (DMR-1420709), NSF grant PHY-1511696, and the Army Research Office-Multidisciplinary Research Initiative under grant W911NF-14-1-0003.

%

\clearpage
\widetext
\setcounter{equation}{0}
\setcounter{figure}{0}
\setcounter{table}{0}
\setcounter{page}{1}
\makeatletter
\renewcommand{\theequation}{S\arabic{equation}}
\renewcommand{\thefigure}{S\arabic{figure}}
\renewcommand{\thetable}{S\arabic{table}}

\noindent \textbf{Materials and Methods}\\

\section{Time-of-flight with focusing}

In order to distinguish the momenta of atoms  with higher resolution, we perform the time-of-flight (TOF) measurement based on the focusing technique \cite{Feng2017}. Our experiment starts with a condensate of cesium atoms in a disk-shaped trap. We pulse on the magnetic field modulation for a short period of $\tau$ to stimulate jet formation. Before the jets start leaving the disk-shaped trap, we quickly tune the scattering length to zero to avoid atomic collisions. We perform time-of-flight by turning off the disk-shaped trap while the harmonic confinement from a cross dipole trap is simultaneously turned on. Atoms with the same momentum will focus to the same location after a quarter of the trapping period, which is 100~ms in our experiment. We carefully tune the dipole trap and optimize the momentum resolution to 0.3~$\hbar k_f$, which later determines the peak width of the rings in our population measurement.

\section{Hamiltonian of five-wave mixing}

Here we derive the Hamiltonian of our system with oscillating atomic interactions induced by magnetic field modulation near a Feshbach resonance. We start with the general form of the Hamiltonian
\begin{equation}
H=\int d^3\bold{r}\Psi^\dagger(\bold r,t)\frac{p^2}{2m}\Psi(\bold r,t)+{g(t)\over 2}\int d^3\bold{r}\Psi^\dagger(\bold r,t)\Psi^\dagger(\bold r,t)\Psi(\bold r,t)\Psi(\bold r,t)+\frac{1}{\mu_0}\int d^3\bold{r}|B(\bold r,t)|^2,
\end{equation}
where $g(t)=4\pi\hbar^2a(t)/m$ is the interaction constant, scattering length $a(t)=a_{dc}+a_{ac}\sin(\omega t)$ is from magnetic field modulation $B(\bold r,t)=B_{ac}\sin(\omega t)$ with $B_{ac}$ the modulation amplitude, and $\mu_0$ is the vacuum permeability. By applying the Fourier transformation
\begin{equation}
\Psi(\bold r,t)={1\over\sqrt V}\sum_k e^{i\bold k \bold r} a_{\bold k},
\end{equation}
where $V$ is the volume of the condensate, we obtain the Hamiltonian in momentum space
\begin{equation}
H=\sum_{\bold k}\epsilon_k a_\vk^\dagger a_\vk+ {g(t)\over 2V}\sum_{\bold k_1,\bold k_2,\Delta\bold k}a^\dagger_{\bold k_1+\Delta \bold k}a^\dagger_{\bold k_2-\Delta \bold k}a_{\bold k_1}a_{\bold k_2}+\frac{1}{\mu_0}\int d^3\bold{r}|B(\bold r,t)|^2.
\end{equation}

When modulating the scattering length using magnetic field within a linear region near a Feshbach resonance, the change of scattering length is proportional to the change of magnetic field $\delta a=\eta\delta B$ with $\eta$ a constant. As we introduce the quantization of the magnetic field, 
\begin{equation}
B = i\sqrt{\hbar\omega\mu_0 \over 2 V} (b-b^\dagger)
\end{equation}
with $b$ ($b^\dagger$) the creation (annihilation) operator for a RF photon, the Hamiltonian can be rewritten as
\begin{eqnarray}
H&=&\sum_{\bold k}\epsilon_k a_\vk^\dagger a_\vk+\hbar\omega b^\dagger b + \sum_{\mathbf{k_1},\mathbf{k_2},\Delta \vk}\left(Aa^\dagger_{\mathbf{k_1}+\Delta \vk} a^\dagger_{\mathbf{k_2}-\Delta \vk} a_{\mathbf{k_1}} a_{\mathbf{k_2}}b+h.c.\right), \label{eq1}
\end{eqnarray}
with coupling constant $A$ defined by
\begin{equation}
A=i{\pi\hbar^2 \eta\over mV}\sqrt{2\hbar\omega\mu_0 \over V}. 
\end{equation}
Notice that we neglect the momentum carried by the RF photon, which is eleven orders of magnitude smaller than $\hbar k_f$ in our experiments. 

After we transfer the operators for atoms and photons into a rotating frame with $a_\vk\rightarrow a_\vk e^{i\epsilon_\vk t/\hbar}$ and $b\rightarrow be^{i\omega t}$, and ignore the fast varying terms, the Hamiltonian becomes
\begin{equation}
H_{int}=\sum_{\mathbf{k_1},\mathbf{k_2},\Delta \vk} (Aa^\dagger_{\vk_1+\Delta \vk} a^\dagger_{\vk_2-\Delta \vk} a_{\vk_1} a_{\vk_2}be^{-i\delta t}+h.c.),
\end{equation}
with $\delta=(\epsilon_{\vk_1+\Delta \vk}+\epsilon_{\vk_2+\Delta \vk}-\epsilon_{\vk_1}-\epsilon_{\vk_2}-\hbar\omega)/\hbar$. When $\delta=0$, the corresponding term in the Hamiltonian is on resonant satisfying energy conservation.

\section{Perturbation theory}\label{sec_perturbation}

\begin{figure}[htbp]
\begin{center}
\includegraphics[width=2.6in]{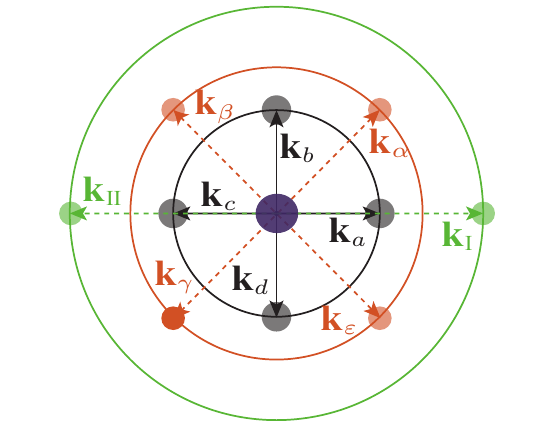}
\caption{Dominant momentum modes after primary and secondary collisions. Here momentum modes $\vk_a$, $\vk_b$, $\vk_c$, and $\vk_d$ in ring 1 are from the primary collisions. Modes $\vk_\alpha$, $\vk_\beta$, $\vk_\gamma$ and $\vk_\epsilon$ in ring 2, and modes $\vk_I$ and $\vk_{II}$ in ring 4 are from secondary collisions.}
\label{sfig1}
\end{center}
\end{figure}

We develop a more quantitative model to describe the initial population growth of atoms in different rings using perturbation theory. For simplicity, we treat the operators $b$ and $b^\dagger$ for RF photons as C-numbers and assume the condensate is far from depletion (Bogoliubov approximation). By only considering the resonant interaction terms, we simplify the Hamiltonian to
\begin{equation}
H_{int}=\hbar\nu\sum_{\vk_1,\vk_2,\Delta \vk}(a^\dagger_{\vk_1+\Delta \vk}a^\dagger_{\vk_2-\Delta \vk} a_{\vk_1} a_{\vk_2}+h.c.),
\end{equation}
with $\nu = {2\pi\hbar a_{ac}\over mV}$. We inspect the equation of motion for $a_{\vk_1+\Delta\vk}$, 
\begin{equation}
i \dot a_{\vk_1+\Delta\vk}=  \nu\sum_{\vk_2,\Delta \vk} a^\dagger_{\vk_2-\Delta\vk}a_{\vk_1}a_{\vk_2},
\end{equation}
where all the energy-and-momentum conserved collisions contribute to the the population growth of mode $\vk_1+\Delta\vk$. However, these collisions do not contribute equally.  Collision processes are more dominant when they involve more macroscopically occupied modes. Here we study the terms that describe dominant collision processes corresponding to our identified pattern $\mathbf{\Phi}$ (Fig.~\ref{sfig1}).

First of all, we consider the primary collisions, the two modes $\vk_a$ and $\vk_c$ are simultaneously occupied due to stimulated inelastic scattering \cite{jets}. The primary interaction terms in Hamiltonian is given by 
 
\begin{equation}
H_P=\hbar\nu a_{a}^{\dagger}a_{c}^{\dagger}a_{0}a_{0}+h.c..
\end{equation}
As a result, the corresponding equations of motion are 

\begin{eqnarray}
i \dot a_{\vk_a}&= &  \nu a^\dagger_{\vk_c}a_{\vk_0}a_{\vk_0},\\
i \dot a_{\vk_c}&= &  \nu a^\dagger_{\vk_a}a_{\vk_0}a_{\vk_0}.
\end{eqnarray}
We solve these equations analytically and obtain
\begin{eqnarray}
a_{a}(t)=a_{a}(0)\cosh(\gamma \tau)-ia_{c}^{\dagger}(0)\sinh(\gamma \tau), \label{s1}\\
a_{c}^{\dagger}(\tau)=a_{c}^{\dagger}(0)\cosh(\gamma \tau)+ia_{a}(0)\sinh(\gamma \tau).\label{s2}
\end{eqnarray}
where we have applied the Bogoliubov approximation $a_0\approx a^\dagger_0\approx \sqrt{N_0}$ and define $\gamma = \nu N_0$ with $N_0$ the total number of atoms in the condensate. Thus the population in ring 1 grows as 

\begin{equation}
\langle a^\dagger_a(\tau) a_a(\tau)\rangle=\langle a^\dagger_c(\tau) a_c(\tau)\rangle=\sinh^2(\gamma \tau), \label{S8}
\end{equation}
assuming that the initial population is zero. Similar solutions also apply to modes $\vk_b$ and $\vk_d$ although they are not correlated to modes $\vk_a$ and $\vk_c$. 

We then proceed to the secondary collisions which involves atoms generated from primary collisions. For ring 2, the dominant interaction terms in the Hamiltonian involving all the eight modes in ring 1 and 2 (Fig.~\ref{sfig1}) are given by 

\begin{eqnarray}
H_{S1}&=&\hbar\nu (a_{\alpha}^{\dagger}a_{d}^{\dagger}a_{a}a_{0}+a_{\epsilon}^{\dagger}a_{b}^{\dagger}a_{a}a_{0}+a_{\beta}^{\dagger}a_{a}^{\dagger}a_{b}a_{0}+a_{\alpha}^{\dagger}a_{c}^{\dagger}a_{a}a_{0} \nonumber \\
& &+a_{\gamma}^{\dagger}a_{b}^{\dagger}a_{c}a_{0}+a_{\text{\ensuremath{\beta}}}^{\dagger}a_{d}^{\dagger}a_{c}a_{0}+a_{\epsilon}^{\dagger}a_{c}^{\dagger}a_{d}a_{0}+a_{\gamma}^{\dagger}a_{a}^{\dagger}a_{d}a_{0}+h.c.).
\end{eqnarray}
For convenience, we first check the equation of motion for $a_\alpha$, which gives
\begin{equation}
i\dot{a}_{\alpha}=\nu(a_{d}^{\dagger}a_{a}\sqrt{N_0}+a_{c}^{\dagger}a_{b}\sqrt{N_0}) \label{s3}
\end{equation}
under Bogoliubov approximation. Assuming that $a_a$, $a_b$, $a_c$ and $a_d$ are unaffected by the secondary collisions, we insert Eq.~(\ref{s1}) and \ref{s2} into \ref{s3}, and get the perturbative solution for population in ring 2,
\begin{equation}
\langle a^\dagger_\alpha (\tau)a_\alpha(\tau)\rangle=\langle a^\dagger_\beta (\tau)a_\beta(\tau)\rangle=\langle a^\dagger_\gamma (\tau)a_\gamma(\tau)\rangle=\langle a^\dagger_\epsilon (\tau)a_\epsilon(\tau)\rangle={1\over2 N_0}\sinh^4(\gamma \tau).\label{S13}
\end{equation}

For the generation of population in ring 4, the relevant interaction terms in the Hamiltonian are given by 
\begin{equation}
H_{S2}=\hbar\nu(a_{I}^{\dagger}a_{0}^{\dagger}a_{a}a_{a}+a_{II}^{\dagger}a_{0}^{\dagger}a_{c}a_{c}+h.c.).
\end{equation}
Following the same procedure, we get the equation of motion for $a_{I}$, 
\begin{equation}
i\dot{a}_{I}=\nu a_{0}^{\dagger}a_{a}^{2}=\nu\sqrt{N_0}a_{a}^{2}.
\end{equation}
Based on the same approach, we obtain
\begin{equation}
\langle a_{I}^{\dagger}a_{I}\rangle=\langle a_{II}^{\dagger}a_{II}\rangle=\frac{1}{8N_0}[\sinh(2\gamma \tau)-2\gamma \tau]^{2}.
\end{equation}

It is easy to find $N_2\propto N^2_1$ from Eqs.~(\ref{S8}) and (\ref{S13}), where $N_1$ is the total population in ring 1 coming from modes like $\vk_a$, $\vk_b$, $\vk_c$ and $\vk_d$, and $N_2$ is the total population in ring 2 resulting from modes like $\vk_\alpha$, $\vk_\beta$, $\vk_\gamma$ and $\vk_\epsilon$. When the driving time $\tau$ is long compared to $1/\gamma$, we can see that
\begin{equation}
\lim_{\tau\rightarrow \infty}{\langle a^\dagger_\alpha (\tau) a_\alpha (\tau)\rangle =\lim_{\tau\rightarrow \infty}  \langle a^\dagger_I (\tau) a_I (\tau) \rangle}= \frac{1}{32N_0}e^{4\gamma \tau}.
\end{equation}
This indicates $N_4\approx N_2\propto N^2_1$ which is consistent with our experimental observations shown in Fig.~2\textbf{B}, where $N_4$ is the total population in ring 4 coming from modes like  $\vk_I$ and $\vk_{II}$.

There are also weaker secondary collisions beside the dominant processes described above. One example is the generation of ring 3 where two atoms from ring 1 in the same angular mode collide. The collision process scatter one atom to ring 3 with momentum $\hbar\sqrt{3}k_f$ and another back to ring 1 at angles of 30$^\circ$ and -60$^\circ$ relative to the initial momentum mode respectively. Due to limited signal level in the population of ring 3, we do not perform quantitative study in this work. Finally, as modes involved in secondary collisions become macroscopically occupied, we anticipate occurrence of tertiary collisions which shall enable generation of even higher harmonics. 

\section{Pattern recognition algorithm}

Our pattern recognition algorithm can be categorized as unsupervised machine learning, which does not require labeled training data. This ensures no human bias to the final recognized common pattern. To identify the key features from the images, the algorithm minimizes a loss function that favors the common pattern by adjusting the orientation of each individual image.

To explain the implementation, we can consider each image a combination of several common patterns that are randomly rotated and contribute to the image with different weights. This randomness results isotropic rings in the average image.  To make the pattern stand out, we align the strongest component while the rest average to a smooth background. This alignment can be achieved by optimizing the angular variance of the average image after rotation. Thus we can define the loss function as the negative of  the angular variance.

\begin{figure}[htbp]
\begin{center}
\includegraphics[width=112mm]{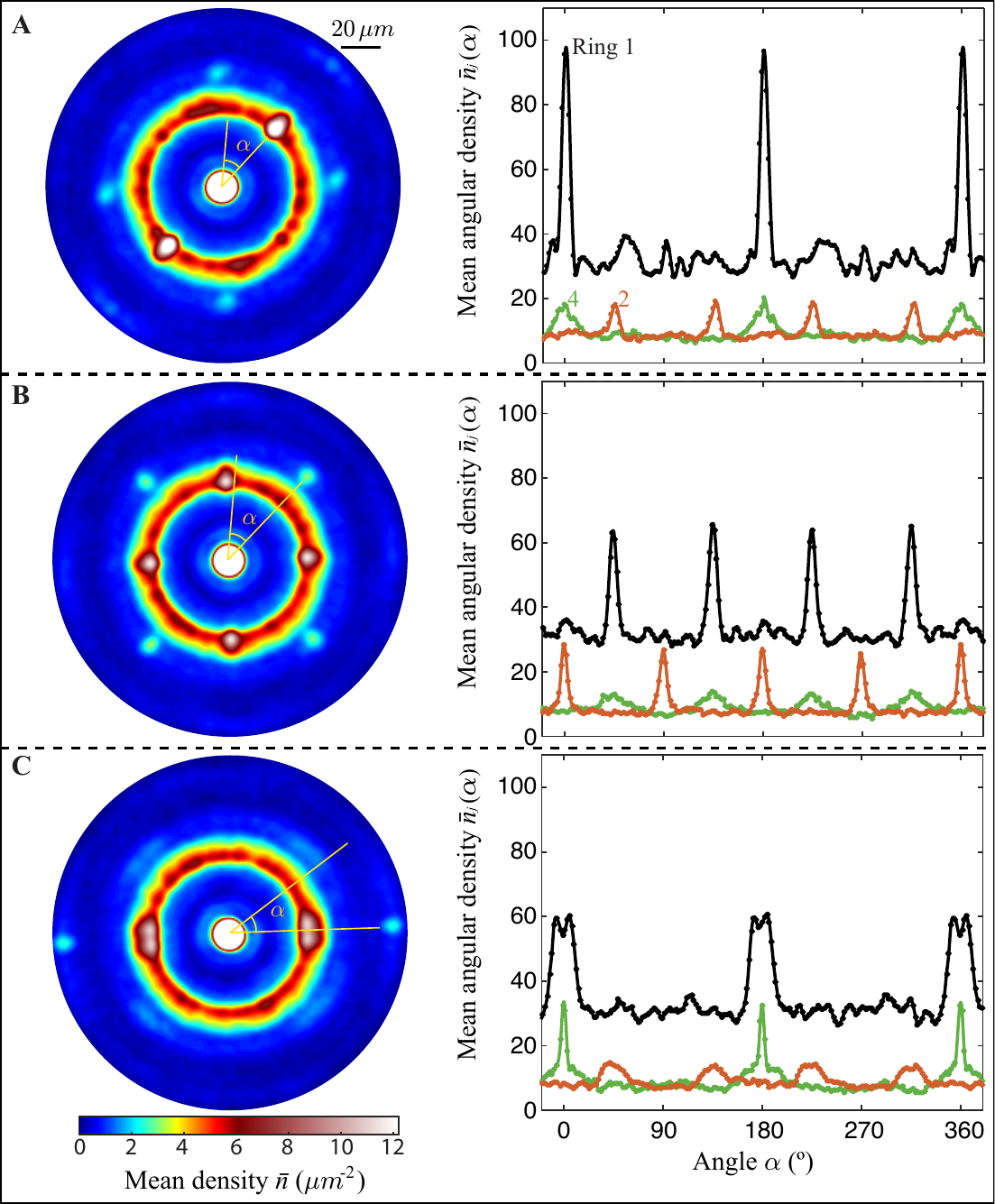}
\caption{Pattern recognition based on individual rings. Left column: the pattern that show up in average image after application of pattern recognition algoritm respectively to ring 1 (\textbf{A}), ring 2 (\textbf{B}) and ring 4 (\textbf{C}). Right column: the corresponding avearge angular density regarding to each ring in the pattern image, where angle $\alpha$ is defined relative to the brightest spot in ring 1, ring 2 and ring 4 respectively.}
\label{figS1}
\end{center}
\end{figure}

We can simplify the calculation of the angular variance by incorporating the rotation symmetry of our system. Since the emitted atoms that contain essential information of the pattern form quantized rings, each image $I_i$ can be faithfully represented by the angular density of rings $\{n_1^{(i)}(\theta),n_2^{(i)}(\theta),n_4^{(i)}(\theta)\}$. Here the angular density of each ring is given by the integral over the radial direction $n_j^{(i)}(\theta) = \int_{R_j-\sigma}^{R_j+\sigma}n^{(i)}(r,\theta)dr$, with $R_j$ the center of the ring and $\sigma$ the $1/e$ width of the ring, which is 7~$\mu m$ in this experiment. Note that we exclude ring 3 due to its low signal level. We then apply a rotation to each individual image with an angle $\theta_i$, thus the angular density becomes $\{n_1^{(i)}(\theta+\theta_i),n_2^{(i)}(\theta+\theta_i),n_4^{(i)}(\theta+\theta_i)\}$. According to our definition, the loss function $L(\{\theta_i\})$ is given by
\begin{equation}
L(\{\theta_i\})=\sum_j^{\{1,2,4\}}L_j(\{\theta_i\}),\\
\end{equation}
where
\begin{equation}
L_j(\{\theta_i\}) = - {1\over 2\pi}\int d\theta \left[{1\over M}\sum_{i=1}^{M} n^{(i)}_j(\theta+\theta_i)-\bar n_j\right]^2
\end{equation}
with
\begin{equation}
\bar n_j={1\over M}\sum_{i=1}^M{1\over 2\pi} \int d\theta  n^{(i)}_j(\theta+\theta_i)
\end{equation}
and $M$~=~209 is the number of images in our dataset. It is easy to see that $\bar n_j$ is a constant independent of $\{\theta_i\}$. We use a derivative-free search algorithm to find local minima of the loss function $L(\{\theta_i\})$. All the local minima yield similar and robust emission pattern $\mathbf{\Phi}$ shown in Fig.~3\textbf{C}.

Beside the results shown in Fig.3 from the main text, we also find similar but different patterns when defining the loss function only based on single ring $j$. The pattern obtained based on ring 1 is very similar to that shown in Fig.~3\textbf{C} (Fig.\ref{figS1}\textbf{A}). The patterns obtained based on ring 2 or ring 4 show much brighter spots in the associated rings. In particular, the angular density $\bar{n}_2$ in Fig.\ref{figS1}\textbf{B} ($\bar{n}_4$ in Fig.\ref{figS1}\textbf{C}) is analogous to the auto-correlations $g^{(2)}_{22}$ ($g^{(2)}_{44}$). Additionally, the angular density $\bar{n}_1$ and $\bar{n}_4$ ($\bar{n}_1$ and $\bar{n}_2$) resemble the cross-correlations $g^{(2)}_{12}$ and $g^{(2)}_{24}$ ($g^{(2)}_{14}$ and $g^{(2)}_{24}$).

\section{Characterization of high-order correlations}

In our experiment, we measure the density distribution in the momentum space and calculate $m$th-order density correlation between $m$ different momentum modes defined as
\begin{equation}
g^{(m)}_{j_1,j_2,...,j_m}(\phi_{12},...,\phi_{1m})=\frac{\langle n_{j_1}(\theta)\prod_{k = 2}^{m}n_{j_k}(\theta+\phi_{1k})\rangle}{\langle n_{j_1}(\theta)\rangle\prod_{k = 2}^{m}\langle n_{j_k}(\theta+\phi_{1k})\rangle},
\end{equation}
where $j_k$~=~1, 2 or 4 is the ring number for the $k$-th mode and $\phi_{1k}$ is the relative angle between 1st mode and $k$-th mode. Here the $\langle.\rangle$ represents angular averaging over $\theta$, followed by ensemble averaging over images. When $g^{(m)}>1$, the $m$ modes $\{n_{j_k}(\theta+\phi_{1k})\}$ are correlated.

High-order correlation, however, may not offer more information about how different modes are correlated with each other than the lower-order correlations. To understand this, we use the third-order correlation function $g^{(3)}_{124}(\phi_{12},\phi_{14})$ as an example. Assuming that $n_2(\theta+\phi_{12})$ and $n_4(\theta+\phi_{14})$ are correlated but neither of them is correlated to $n_1(\theta)$, we have
\begin{equation}
\langle n_1(\theta) n_2(\theta+\phi_{12}) n_4(\theta+\phi_{14})\rangle=\langle n_1(\theta)\rangle \langle n_2(\theta+\phi_{12}) n_4(\theta+\phi_{14})\rangle.
\end{equation}
Consequently, the third-order correlation reduces to the second-order correlation, 
\begin{equation}
g^{(3)}_{124}(\phi_{12},\phi_{14})={\langle n_1(\theta)\rangle \langle n_2(\theta+\phi_{12}) n_4(\theta+\phi_{14})\rangle \over \langle n_1(\theta)\rangle \langle n_2(\theta+\phi_{12}) \rangle\langle n_4(\theta+\phi_{14})\rangle}={\langle n_2(\theta+\phi_{12}) n_4(\theta+\phi_{14})\rangle \over  \langle n_2(\theta+\phi_{12}) \rangle\langle n_4(\theta+\phi_{14})\rangle}=g^{(2)}_{24}(\phi_{12}-\phi_{14}). \label{S35}
\end{equation}

In order to obtain how genuinely $m$ modes are correlated to each other, we extract the connected correlation function $\tilde{g}^{(m)}$ by substracting the contributions from all the lower-order correlations,
\begin{equation}
\tilde{g}^{(m)}=g^{(m)}-g^{(m)}_{dis}, 
\end{equation}
where $g^{(m)}_{dis}$ is the disconnected part from lower-order correlations. According to Wick's decomposition \cite{Schweigler2017}, $g^{m}_{dis}$  is given by
\begin{equation}
g^{(m)}_{dis}={1\over \prod_{k=1} \langle n_{j_k}(\theta+\phi_{1k})\rangle}\sum_\Lambda \left[ (N_\Lambda-1)! (-1)^{N_\Lambda-1}\prod_{B\in \Lambda} \langle\prod_{k \in B} n_{j_k}(\theta+\phi_{1k})\rangle  \right].
\end{equation}
Here the sum $\sum_\Lambda$ runs over all possible partitions $\Lambda$ of $\{1,2,\ldots,m\}$, the first product in the square brackets runs over all blocks $B$ of the partition and the second product runs over all elements $k$ in the block; $N_\Lambda$ is the number of blocks in the partition. We have absorbed $n_{j_1}(\theta)$ into the product with $\phi_{11}$~=~0. 

Take $g^{(3)}_{dis}(\phi_{12},\phi_{14})$ as an example again, the disconnected part is
\begin{eqnarray}
g^{(3)}_{dis}(\phi_{12},\phi_{14})&=&g^{(1)}_1g^{(2)}_{24}(\phi_{12}-\phi_{14})+g^{(1)}_2g^{(2)}(\phi_{14})+g^{(1)}_4g^{(2)}_{12}(\phi_{12})-2g^{(1)}_1g^{(1)}_2g^{(1)}_4 \nonumber \\
&=&g^{(2)}_{24}(\phi_{12}-\phi_{14})+g^{(2)}(\phi_{14})+g^{(2)}_{12}(\phi_{12})-2.
\end{eqnarray}
As an example, when we evaluate the connected part of the third-order correlation in Eq.~(\ref{S35}), we obtain
\begin{eqnarray}
\tilde{g}^{(3)}_{124}(\phi_{12},\phi_{14})&=&g^{(3)}_{124}(\phi_{12},\phi_{14})-g^{(3)}_{124,dis}(\phi_{12},\phi_{14})=0,
\end{eqnarray}
which shows no genuine high-order correlation as we expect. However, in contrast to this trivial example, our experiment shows significant non-trivial third-order and eighth-order correlations.

Beside experimental characterization, we can also calculate the correlations between different momentum modes based on the perturbation theory in section \ref{sec_perturbation}. Such as for ring 1 and ring 2, there are
\begin{eqnarray}
\langle a_{a}^{\dagger}(\tau)a_{\alpha}^{\dagger}(\tau)a_{\alpha}(\tau)a_{a}(\tau)\rangle&=&{1\over N_0}\sinh^6(\gamma \tau),\\
\langle a_{c}^{\dagger}(\tau)a_{\alpha}^{\dagger}(\tau)a_{\alpha}(\tau)a_{c}(\tau)\rangle&=&{1\over2 N_0}\sinh^2(\gamma \tau)[\sinh^2(\gamma \tau)+\cosh^2(\gamma \tau)].
\end{eqnarray}
Thus the normalized second-order cross correlations at 45$^\circ$ and 135$^\circ$ between ring 1 and 2 are
\begin{eqnarray}
g^{(2)}_{12}(\phi=45^\circ)&=&\frac{\langle a_{1}^{\dagger}(\tau)a_{\alpha}^{\dagger}(\tau)a_{\alpha}(\tau)a_{1}(\tau)\rangle}{\langle a_{\alpha}^{\dagger}(\tau)a_{\alpha}(\tau)\rangle\langle a_{1}^{\dagger}(\tau)a_{1}(\tau)\rangle}=2, \\
g^{(2)}_{12}(\phi=135^\circ)&=&\frac{\langle a_{3}^{\dagger}(\tau)a_{\alpha}^{\dagger}(\tau)a_{\alpha}(t)a_{3}(\tau)\rangle}{\langle a_{\alpha}^{\dagger}(\tau)a_{\alpha}(\tau)\rangle\langle a_{3}^{\dagger}(\tau)a_{3}(\tau)\rangle}=1+(\coth(\gamma \tau))^{2}\xlongrightarrow[\tau\rightarrow\infty] {}  2.
\end{eqnarray}
For ring 1 and 4, we expect
\begin{eqnarray}
\langle a_{1}^{\dagger}(\tau)a_{I}^{\dagger}(\tau)a_{I}(\tau)a_{1}(\tau)\rangle&=&{3\over 8N_0}\sinh^2(\gamma \tau)\left[\sinh(2\gamma \tau)-2\gamma \tau\right]^2,\\
\langle a_{3}^{\dagger}(\tau)a_{I}^{\dagger}(\tau)a_{I}(\tau)a_{3}(\tau)\rangle&=&{1\over 8N_0}\sinh^2(\gamma \tau)\left\{\left[\sinh(2\gamma \tau)-2\gamma \tau\right]^2+8\sinh^4(\gamma \tau) \right\}.
\end{eqnarray}
Then the normalized second-order cross correlations at 0$^\circ$ and 180$^\circ$ between ring 1 and 4 are
\begin{eqnarray}
g^{(2)}_{14}(\phi=0^\circ)&=&\frac{\langle a_{1}^{\dagger}(\tau)a_{I}^{\dagger}(\tau)a_{I}(\tau)a_{1}(\tau)\rangle}{\langle a_{I}^{\dagger}(\tau)a_{I}(\tau)\rangle\langle a_{1}^{\dagger}(\tau)a_{1}(\tau)\rangle}=3, \\
g^{(2)}_{14}(\phi=180^\circ)&=&\frac{\langle a_{3}^{\dagger}(\tau)a_{I}^{\dagger}(\tau)a_{I}(\tau)a_{3}(\tau)\rangle}{\langle a_{I}^{\dagger}(\tau)a_{I}(\tau)\rangle\langle a_{3}^{\dagger}(\tau)a_{3}(\tau)\rangle}=1+8{ \sinh^4(\gamma \tau)\over\left[\sinh(2\gamma \tau)-2\gamma \tau\right]^2 }\xlongrightarrow[\tau\rightarrow\infty] {}  3.
\end{eqnarray}
For ring 2 and ring 4, we expect
\begin{eqnarray}
\langle a_{\alpha}^{\dagger}(t)a_{I}^{\dagger}(\tau)a_{I}(\tau)a_{\alpha}(\tau)\rangle&=& {3\over 8 N_0^2}\sinh^4(\gamma \tau)\left[\sinh(2\gamma \tau)-2\gamma \tau\right]^2, \\
\langle a_{\alpha}^{\dagger}(\tau)a_{II}^{\dagger}(\tau)a_{II}(\tau)a_{\alpha}(\tau)\rangle&=&{1\over 8 N_0^2}\sinh^4(\gamma \tau)\left\{\left[\sinh(2\gamma \tau)-2\gamma \tau\right]^2+8\sinh^4(\gamma \tau)\right\} \nonumber.
\end{eqnarray}
The second-order correlations between ring 2 and 4 are
\begin{eqnarray}
g^{(2)}_{24}(\phi=45^\circ)&=&\frac{\langle a_{\alpha}^{\dagger}(\tau)a_{I}^{\dagger}(\tau)a_{I}(\tau)a_{\alpha}(\tau)\rangle}{\langle a_{I}^{\dagger}(\tau)a_{I}(\tau)\rangle\langle a_{\alpha}^{\dagger}(\tau)a_{\alpha}(\tau)\rangle}=3, \\
g^{(2)}_{24}(\phi=135^\circ)&=&\frac{\langle a_{\alpha}^{\dagger}(\tau)a_{III}^{\dagger}(\tau)a_{III}(\tau)a_{3}(\tau)\rangle}{\langle a_{III}^{\dagger}(\tau)a_{III}(\tau)\rangle\langle a_{\alpha}^{\dagger}(\tau)a_{\alpha}(\tau)\rangle}=1+8{ \sinh^4(\gamma \tau)\over\left[\sinh(2\gamma \tau)-2\gamma \tau\right]^2 }\xlongrightarrow[\tau\rightarrow\infty] {}  3.
\end{eqnarray}

In addition, we also inspect $g_{124}^{(3)}$ and $\tilde g_{124}^{(3)}$ for mode $\vk_a$, $\vk_\alpha$ and $\vk_I$. According to the perturbation theory, we have
\begin{equation}
\langle a^\dagger_{1} (\tau)a^\dagger_{\alpha} (\tau)a^\dagger_{I} (\tau)a_{I}(\tau)a_{\alpha}(\tau)a_{1}(\tau)\rangle={3\over 4N_0^2}\sinh^6(\gamma \tau)\left[\sinh(2\gamma \tau)-2\gamma \tau\right]^2.
\end{equation}
Thus $g_{124}^{(3)}(\phi_{12} = 45^\circ,\phi_{14} = 0^\circ)~=~12$, and $\tilde g_{124}^{(3)}$ is given by
\begin{eqnarray}
\tilde g_{124}^{(3)}(\phi_{12} = 45^\circ,\phi_{14} = 0^\circ)&=&g_{124}^{(3)}(\phi_{12} = 45^\circ,\phi_{14} = 0^\circ)-\\
& &g_{12}^{(2)}(\phi_{12} = 45^\circ)-g_{14}^{(2)}(\phi_{12} = 0^\circ)-g_{24}^{(2)}(\phi_{24} = 45^\circ)+2 \nonumber\\
&=&6 \nonumber.
\end{eqnarray}
The non-zero connected part of third-order correlation represents genuine three-body correlations. The same method applies to even higher-order correlation functions. Note that the theoretical values of these correlations are generally large than those observed in experiments. This is mainly due to the significant depletion of the condensates which is not accounted for in the perturbation theory.

\section{Effective three-body collisions}

We inspect the connected third-order correlation function $\tilde g^{(3)}_{124}$ and find weak correlations along lines across the peaks in Fig.4\textbf{B} in the main text. We attribute these lines to effective three-body collisions where the atoms interact with the modulation field twice through an intermediate state, similar to the Raman transitions in quantum optics. To see this, we derive the effective Hamiltonian that describes this process. 

\begin{figure}[htbp]
\begin{center}
\includegraphics[width=168mm]{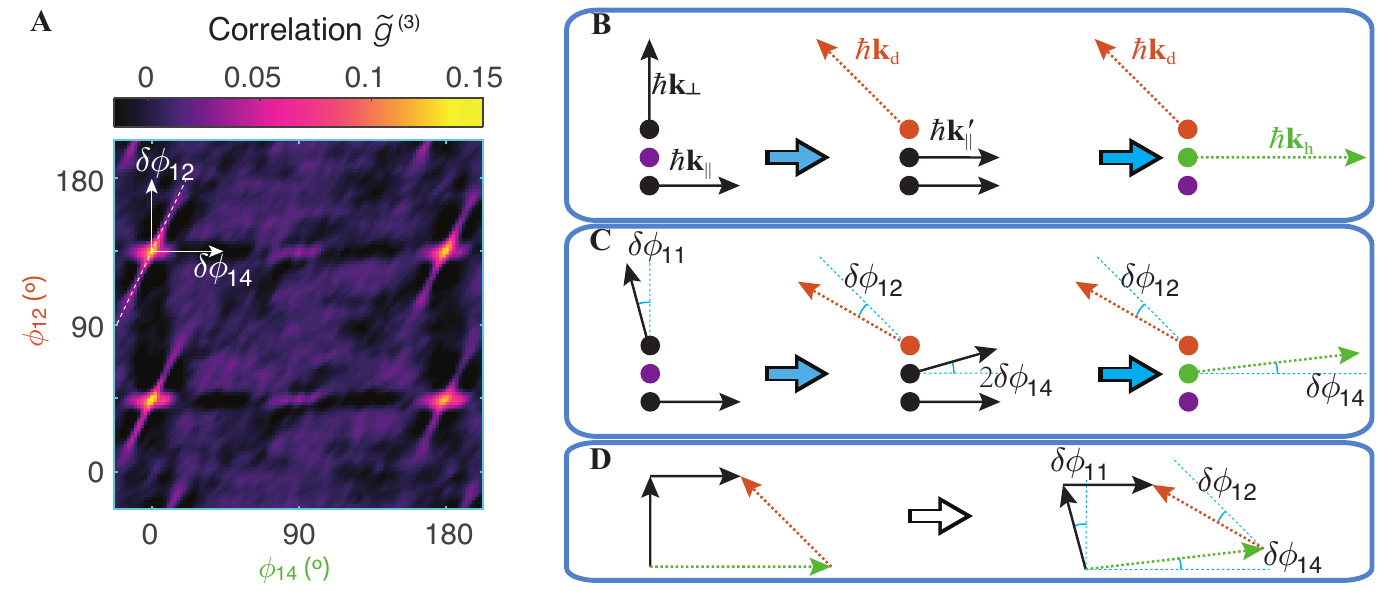}
\caption{Thin correlation lines in third-order correlations $\tilde g_{124}^{(3)}$. \textbf{A} shows thin lines across the peaks in the connected third-order correlation $\tilde g_{124}^{(3)}$ from our measurement. One example is illustrated with white dashed line across the peak position $(\phi_{14},\phi_{12})$~=~$(0^\circ,135^\circ)$. The angle deviations are defined as $\delta\phi_{14}$ and $\delta\phi_{12}$, respectively. \textbf{B} illustrates an exemplary two-step collision process that generates a pair of atoms in ring 2 and 4 from two atoms in ring 2 in single-photon resonant way. Here $|\vk_\perp|=|\vk_\parallel|=|\vk_\parallel^\prime|=k_f$, $|\vk_d|=\sqrt{2}k_f$ and $|\vk_h|=2k_f$. Each step conserves energy and momentum. \textbf{C} presents the Raman process responsible for the thin line, where each step is slightly off resonant but the whole process conserves energy. Here each step still satisfies momentum conservation and $\delta\phi_{11}$ is the angle deviation of $\vk_\perp$ from its resonant orientation. \textbf{D} shows the phase-matching condition for the  effective three-body collision.}
\label{sfig2}
\end{center}
\end{figure}

Following the derivation in section \ref{sec_perturbation}, the general form of the Hamiltonian is 
\begin{equation}
H_{int}(t)=\hbar\nu\sum (a^\dagger_{\vk_1+\Delta k} a^\dagger_{\vk_2-\Delta k} a_{\vk_1} a_{\vk_2}e^{-i\delta t}+h.c.).
\end{equation}
When $\delta\neq0$, the corresponding term in the Hamiltonian is off resonant and such direct collision is not allowed. However, two such off-resonant terms can potentially cancel the detuning together to yield a resonant coupling in a time averaged Hamiltonian.

To perform this time averaging,  we first look at the time evolution operator $U(t)$ that satisfies
\begin{eqnarray}
i\hbar {\partial U(t)\over\partial t}&=&H(t)U(t).
\end{eqnarray}
When $H(t)$ is changing rapidly in time, we can average the overall evolution operator and eliminate the rapid oscillating terms to gain physics at slow time scale. Thus we define a time average function $F(t)$ with peaks at $t=0$ and $\int dt F(t)=1$ that spans over a short period of time. The detail form of $F(t)$ is not important. Therefore the average evolution operator $\overline{ U(t)}$ is $\overline{ U(t)}=\int dt' F(t-t') U(t')$ and the equation of motion becomes

\begin{equation}
i\hbar{\partial\over \partial t}\overline{U(t)}=\overline{H(t)U(t)}.
\end{equation}
Effectively, we expect the equation of motion to be $i\hbar{\partial\over \partial t}\overline{U(t)}={H_{eff}(t)}\overline{U(t)}.$ As a result, the general form of the effective Hamiltonian after time averaging is 
\begin{equation}
H_{eff}(t)=\overline{H(t)U(t)}\left[\overline {U(t)}\right]^{-1}.
\end{equation}
By expanding this effective Hamiltonian only to the first order, we have the result of 
\begin{equation}
H_{eff}(t)=\overline{H(t)}+{1\over 2}\left(\overline{\left[H(t),U_1(t)\right]}-\left[\overline{H(t)},\overline{U_1(t)}\right]\right),\label{S56}
\end{equation}
with $U_1(t)={1\over i\hbar}\int^t_0 dt' H(t')$.

As an particular example shown in Fig. \ref{sfig2}, we are interested in the line across the peak at $(\phi_{14},\phi_{12})$~=~$(0^\circ,135^\circ)$. Here the angle deviations from the peak position are defined as $\delta\phi_{14}$ and $\delta\phi_{12}$ respectively. The line gives $\delta\phi_{12}\approx2\delta\phi_{14}$ that is universally true close to every peak. The two responsible secondary collisions are shown in Fig.\ref{sfig2}\textbf{B}. When both of the collisions are on single-photon resonance, they can happen simutaneously and directly contribute to the peak. When individual collision is off-resonant, they have to happen in a sequential manner to form resonant Raman coupling shown in Fig.\ref{sfig2}\textbf{C}.     

Particularly, the Hamiltonian involving these momentum modes is written as 
\begin{equation}
H_{int}=\hbar\nu(e^{-i\delta t} a^\dagger_{\vk_d}a^\dagger_{\vk'_\parallel}a_{\vk_\perp}a_0+e^{i\delta t}a^\dagger_{\vk_h}a^\dagger_0a_{\vk_\parallel}a_{\vk'_\parallel} +h.c).
\end{equation}
Plugging this in to Eq.~(\ref{S56}), we get the effective Hamiltonian,
\begin{equation}
H_{eff}=-\hbar{\nu^2\over  \delta}(a^\dagger_{\vk_h} a^\dagger_{ \vk_d}a^\dagger_0 a_0 a_{\vk_\perp}a_{\vk_\parallel}+h.c).
\end{equation}
which shows an effective three-body interaction. To explain in detail in Fig.\ref{sfig2}\textbf{C}, we can assume small perturbations so that $|\vk_\perp|$~=~$(1-\varepsilon)k_f$ with angle deviation of $\delta\phi_{11}$, $\vk_d$ deviate from its original angle by $\delta\phi_{12}$ with the length of the vector unperturbed, and $\vk_h$ deviate from its original angle by $\delta\phi_{14}$. For simplicity, we also assume that the intermediate $|\vk_\parallel^\prime|=k_f$, therefore $\vk_\parallel^\prime$ deviates from the horizontal direction by $2\delta\phi_{14}$ and $|\vk_h|=2\cos(\delta\phi_{14})k_f$ as a result of momentum conservation in the second step. Due to total energy conservation for the two-step process, the detuning for the first step $\delta_f = \frac{\hbar^2k_f^2}{2m}[1-(1-\varepsilon)^2]$ should cancel the detuning for the second step $\delta_s = -\frac{2\hbar^2k_f^2}{m}[1-\cos^2(\delta\phi_{14})]$. Thus we have 
\begin{equation}
\cos(2\delta\phi_{14})=\frac{1}{2}[1+(1-\varepsilon)^2].\label{eqstep1}
\end{equation}
Further considering the momentum conservation in the first step, we obtain two equations from horizontal and vertical component respectively, 
\begin{eqnarray}
-(1-\varepsilon)\sin(\delta\phi_{11})&=&\sqrt{2}\cos(135^\circ+\delta\phi_{12})+\cos(2\delta\phi_{14})\label{eqstep21}\\
(1-\varepsilon)\cos(\delta\phi_{11})&=&\sqrt{2}\sin(135^\circ+\delta\phi_{12})+\sin(2\delta\phi_{14}).\label{eqstep22}
\end{eqnarray}
As a result, we have 
\begin{equation}
\delta\phi_{12} = 2\delta\phi_{14}+\arcsin[\frac{2-\cos(2\delta\phi_{14})}{\sqrt{2}}].
\end{equation} 
In the perturbation regime,  $\delta\phi_{12}\approx2\delta\phi_{14} $ agrees with the line in the third-order correlation function in Fig.4\textbf{B}.

\end{document}